\documentstyle[12pt]{article}
\newcommand{\beq}{\begin{equation}}
\newcommand{\eeq}[1]{\label{#1}\end{equation}}
\newcommand{\beqn}{\begin{eqnarray}}
\newcommand{\eeqn}[1]{\label{#1}\end{eqnarray}}
\newcommand{\ba}{\begin{array}}
\newcommand{\ea}{\end{array}}

\newcommand{\Bs}{\not\!\! B}
\newcommand{\ds}{\!\!\not\!\partial}
\newcommand{\as}{\not\!\! a}

\newcommand{\D}{{\cal{D}}}

\newcommand{\Z}{{\cal Z}}

\newcommand{\bp}{\psi^{\dagger}}
\newcommand{\hp}{\hat{\psi}}

\def\CVO#1#2#3{\!\left( \matrix{ #1 \cr #2 \ #3 \cr} \right)\!}

\begin{document}

%

\vspace*{-2 cm}
\hfill\vbox{
\hbox{~~}
\hbox{\it La Plata-Th 98/05}
\hbox{\it CCNY-HEP 98/2}
\hbox{hep-th/9803047}
}
\vspace{1cm}

\centerline{\Large \bf Quasiparticle operators with non-Abelian}
\centerline{\Large \bf braiding statistics}

\vspace{.6 cm}

\centerline{\large \bf
Daniel C. Cabra$^a$\footnote{CONICET, Argentina. E-mail address:
cabra@venus.fisica.unlp.edu.ar}, \,
Enrique F. Moreno$^b$\footnote{CUNY, New York.
E-mail address: moreno@scisun.sci.ccny.cuny.edu}}

\centerline{\bf and}

\centerline{\large \bf
Gerardo L. Rossini$^c$\footnote{On leave of absence from
Universidad de La Plata, CONICET, Argentina.
E-mail address: rossini@venus.fisica.unlp.edu.ar}}

\vspace{.6cm}

\centerline{\small\it $^a$Departamento de F\'{\i}sica,
Universidad Nacional de La Plata,}
\centerline{\small\it C.C. 67 (1900) La Plata, Argentina}
\centerline{\small\it Facultad de Ingenier\'\i a, Universidad Nacional de 
Lomas de Zamora,}
\centerline{\small\it Cno. de Cintura y Juan XXIII, (1832), Lomas de Zamora, 
Argentina}
\vspace{0.2cm}
\centerline{\small\it $^b$ Physics Department,
City College of the City University of New York}
\centerline{\small\it New York NY 10031, USA}
\centerline{\small\it Physics Department, Baruch College, The City University of
New York}
\centerline{\small\it New York NY 10010, USA}
\vspace{0.2cm}
\centerline{\small\it $^c$ Center for Theoretical Physics,}
\centerline{\small\it Laboratory for Nuclear Science and Department of Physics,}
\centerline{\small\it Massachusetts Institute of Technology,}
\centerline{\small\it Cambridge, MA 02139, USA}

\vspace{1.5 cm}

\centerline{\bf Abstract}
\vspace{0.1cm}
{We study the gauge invariant fermions in the fermion coset representation
of $SU(N)_k$ Wess-Zumino-Witten models which create, by construction, the
physical excitations (quasiparticles) of the theory. We show that they
provide an explicit holomorphic factorization of $SU(N)_k$ Wess-Zumino-Witten
primaries and  satisfy non-Abelian braiding relations.}

\newpage

\noindent {\it i) Introduction}

\vspace{.5cm}

\setcounter{footnote}{0}

The notion of generalized statistics has attracted a great deal of attention
in the last 15 years (see {\it e.g.}\ \cite{Wil,Lei} and references therein),
both in $1+1$ and in $2+1$ dimensional theories.

The most natural generalization of the traditional classification of
particles
into bosons and fermions corresponds to the possibility of having an arbitrary
phase when two particles are interchanged. This is allowed in $2+1$ dimensions
since the statistics is characterized by the Braid group, $B_n$,
(instead of $S_n$ as it is in higher dimensions), which admits more general
representations.

A further generalization is to consider the case of higher dimensional
(non-Abelian) representations of $B_n$ which would correspond to
non-Abelian statistics.
In fact, the concept of non-Abelian statistics was introduced
in the context of the
Fractional Quantum Hall Effect (FQHE) in \cite{MR} as the statistics
of the quasiparticles of the so-called Pfaffian state, (see also \cite{BW},
where more general examples were treated).
The approach in these papers is based
on the construction of trial ground state wave functions of FQHE systems
using
the Conformal Blocks (CB) of a Conformal Field Theory, and the statistics
appears as a consequence of the braiding properties of these CB's, which are
known \cite{MS,AGS}.
In particular, in \cite{BW} the mentioned CB's correspond to
those of the $SU(N)_k$ Wess-Zumino-Witten (WZW) theories \cite{Wi}.

In the present Letter we address this issue from a different point of view:
we aim to construct a concrete realization of quasiparticle operators
satisfying non-Abelian braiding relations.

To this end we first show that the holomorphic and anti-holomorphic factors
of the WZW primary fields in a $SU(N)_k$ WZW theory can be constructed
using its fermionic coset description. In fact, the natural observables in
the
fermionic description, which are the gauge invariant fermions (GIF's)
\cite{CR1,CM,CR},
correspond to the holomorphic factors of the WZW primaries.

Finally, by evaluating the Operator Product Algebra (OPA) of the
GIF's,
we show that they satisfy non-Abelian braiding relations \cite{TK,AGS},
thus giving the desired explicit operator realization
of quasiparticle operators with non-Abelian braiding.

The present approach could be useful in the context of the FQHE for systems
whose quasiparticles obey non-Abelian statistics.

As for the holomorphic factorization that we present, it corresponds to a
very interesting issue
from a more formal point of view, and has been the subject of recent
investigations \cite{HO}.

The present approach could also be useful in connection
with the spinon construction developed in \cite{BLS} and
with the construction of quasiparticle representations
of the characters for Conformal Field Theories \cite{MCoy}.

\vspace{.5cm}

\noindent {\it ii) $SU(N)_k$ Wess-Zumino-Witten (WZW) theory as a
fermionic coset}

\vspace{.5cm}

To make the paper self-consistent and set our conventions we will
first recall the fermionic coset representation of the $SU(N)_k$
WZW theory \cite{NS}. The action is given by \beq S=\frac{1}{\sqrt
2\pi} \int d^2 x {\bar \psi}^{i \alpha}\left( (i \ds +\as
)\delta_{ij}\delta_{\alpha\beta}+ \delta_{ij}\Bs_{\alpha\beta}
\right) \psi^{j\beta} , \eeq{1} where the fermions
$\psi^{i\alpha}$ ($i=1,\cdots ,N$, $\alpha =1,\cdots ,k$) are in
the fundamental representation of $U(Nk)$ and the $U(1)$ gauge
field $a_{\mu}$ and the $SU(k)$ gauge field $B_{\mu}$ act as
Lagrange multipliers implementing the constraints 

\beq
j_{\mu}|{\it phys}>=0 ,~~~~~~~~J_{\mu}^a|{\it phys}>=0 ~~~~ 
(a=1,\cdots , k^2-1)~,
\eeq{2} 
for the $U(1)$ and the $SU(k)$ currents respectively. This corresponds
to the identification: \beq SU(N)_k \equiv \frac{U(Nk)}{SU(k)_N
\times U(1)} , \eeq{3} which is understood as an equivalence
between the correlation functions of corresponding fields in the
two theories.

The fundamental field $g$ of the
bosonic $SU(N)_k$
WZW theory is represented in terms of fermions by the bosonization
formula
\footnote{Our conventions are $\psi=\left(\begin{array}{l}
\psi_L \\
\psi_R
\end{array}\right)$ and $\gamma_i$ are the Pauli matrices,
$\gamma_1=\left(\begin{array}{ll}
0 & 1 \\
1 & 0
\end{array}\right)$ and
$\gamma_2=\left(\begin{array}{rr}
0 & -i \\
i & 0
\end{array}\right)$.}
\cite{NS}

\beq g^{ij}=\psi_L^{i\dagger}\psi_R^{j} \eeq{4} where the $SU(k)$
indices are summed out. Their conformal dimensions are given by
$(h_g, \bar h_g)=(\frac{N^2-1}{2N(k+N)}, \frac{N^2-1}{2N(k+N)}).$

Other integrable representations are constructed as appropriately symmetrized
products of these fundamental fields.
In the particular $N=2$ case, higher spin integrable representations are
constructed as
\beqn
g^{(j)~i_1,...,i_{2j}}_{j_1,...,j_{2j}}&=&{\cal S} \left(:g^{i_1}_{j_1}
...g^{i_{2j}}_{j_{2j}}:\right)
\nonumber \\
&=&{\cal S} \left(:\psi_L^{i_1\dagger}\psi_R^{j_1}
...\psi_L^{i_{2j}\dagger}\psi_R^{j_{2j}}:\right),
\eeqn{5}
where ${\cal S}$ stands for symmetrization over
the left and right indices separately 
and $j$ takes the values $j=0,1/2,1,....,k/2$.
This restriction in the spin of the
representation has its origin in the selection rules imposed by
the affine (Kac-Moody) symmetry \cite{GW,FGK}.
It is interesting to note that in the fermion description of $SU(N)_k$
the presence of a second index $\alpha$ in
the fermion fields $\psi^{i\alpha}$, running from $1$ to $k$, allows
for the construction of symmetrized products of at most $k$ bilinears.
In this way we obtain only the allowed integrable representations,
other representations being forbidden by the Pauli principle.

It will be useful for later purposes to review the decoupled picture.
The partition function is given by
\beq
\Z=\int \D\bp \D\psi \D a_{\mu} \D B_{\mu} \exp(-S),
\eeq{a.1}
where $S$ is given in eq.\ (\ref{1}).

The decoupling transformations are\footnote{Further conventions are:
$z=x_1+i x_2$, ${\bar z}=x_1-i x_2$,
$\partial \equiv \frac{\partial}{\partial z}$,
$\bar{\partial} \equiv \frac{\partial}{\partial \bar z}$,
$a=(a_1-ia_2)/2$, $\bar{a}=(a_1+ia_2)/2$,
$B=(B_1-iB_2)/2$, $\bar{B}=(B_1+iB_2)/2$
and
$h=\exp(\phi +i\eta)$,
$\bar h=\exp(-\phi + i\eta)$.}

\beq
\begin{array}{ll}
\psi_L=h V^{-1}\chi_L & \psi_R={\bar h} U^{-1} \chi_R,\\
a=i {\bar h}  \partial {\bar h}^{-1} &
\bar{a}= i h {\bar \partial} h^{-1},  \\
B=i U^{-1} \partial U &
\bar{B}= i V^{-1} {\bar \partial} V.  \\
\end{array}
\eeq{a.2}

Taking into account the gauge fixing procedure and the Jacobians
associated to the change of variables above \cite{FiPol} one arrives at
the desired decoupled form for the partition function:
\beq
\Z=\Z_{ff}\Z_{fb}\Z_{WZW}\Z_{gh},
\eeq{a.3}
where
\beqn
\Z_{ff} &=& \int \D{\bar \chi} \D\chi \exp( -\frac{1}{\pi}
\int(\chi^{i\alpha\dagger}_L \bar{\partial} \chi_L^{i\alpha} +
\chi^{i\alpha\dagger}_R \partial \chi_R^{i\alpha})d^2x), \nonumber \\
\Z_{fb} &=& \int \D\phi
\exp(\frac{Nk}{2\pi}\int \phi \Delta \phi d^2x),
\nonumber
\\
\Z_{WZW} &=& \int \D {\tilde g} \exp\left((2k+N)\Gamma [{\tilde g}]\right).
\eeqn{a.4}
$-(2k+N) \Gamma [{\tilde g}]$ is the level $-(2k+N)$ WZW action
\cite{Wi} for the
gauge invariant combination ${\tilde g} = V {U^{-1}}$,
and $Z_{gh}$ corresponds to the Fadeev-Popov ghosts partition
function, whose explicit form will not be needed.

In particular,  the central charge is easily evaluated as the sum of four
independent
contributions coming from the different sectors, $c_{ff}= N k$,
$c_{fb}=1$, $c_{WZW}=(2k+N)(k^2-1)/(k+N)$ and $c_{gh}=-2 k^2$,
thus giving
\beq
c=\frac{k(N^2-1)}{k+N} ,
\eeq{a4}
which corresponds to the central charge of the $SU(N)_k$ WZW
action.

\vspace{.5cm}

\noindent {\it iii) Gauge invariant fermions and holomorphic
factorization}

\vspace{.5cm}
The coset theory, defined by the Lagrangian (1), is manifestly
invariant under gauge transformations $U(1)\times SU(k)$. Consequently,
the original fermion fields $\psi$ are not ``physical" operators, as
they do not commute with the associated BRST charges. It is then natural to find
a new set of
fermionic variables invariant under the BRST symmetry
that, by construction, create the physical excitations of the theory.

For the fermionic coset theory there is a natural candidate for this
gauge invariant fermion field. The gauge degrees of freedom
$a_{\mu}$ and  $B_{\mu}$ that enter in the Lagrangian (1) are of
topological nature since they do not have kinetic terms. In fact, their
equations of motion are just the flat connection conditions
$F_{\mu \nu}^a[B]=F_{\mu \nu}[a]=0$. Then, we can attach to the
fermion an infinite flux line that, due to the zero-curvature
condition, is only dependent on the end point, guaranteeing its
locality. This infinite Wilson line absorbs the gauge variation of
the fermion field and then the compound object has the desired
properties. We will show later that, in addition, this GIF's are
also chiral and can be identified with the vertex operators
of the coset model in the sense defined in \cite{TK}.

Let us start defining the gauge invariant fermion fields \cite{CR1}:
\beq
\ba{l}
\hp^{i\alpha}(x)=e^{i\int_x^{\infty} dz^{\mu}a_{\mu}}
P\left(e^{i\int_x^{\infty} dz^{\mu}B_{\mu}}\right)^{\alpha\beta}
\psi^{i\beta}(x)\\
\hp^{i\alpha\dagger}(x)=\psi^{j\beta\dagger}(x)
P\left(e^{i\int_x^{\infty} dz^{\mu}B_{\mu}}\right)^{\beta\alpha\dagger}
e^{-i\int_x^{\infty} dz^{\mu}a_{\mu}}
\ea
\eeq{6}

As we stated above, the Schwinger line integrals in (\ref{6}) do not
depend on the choice of the path due to the zero curvature condition
satisfied by the gauge connections $a_{\mu}$ and $B_{\mu}$.

In order to analyze the properties of the GIF's it is useful to
work with decoupled variables (\ref{a.2}), where
things are more easily tractable.

In terms of these variables, the gauge invariant fermions are given by
\beq
\hp^{i\alpha}_L(x)=e^{i\int_x^{\infty} dz^{\mu}a_{\mu}}
P\left(e^{i\int_x^{\infty} dz^{\mu}B_{\mu}}\right)^{\alpha\beta}
h (V^{-1})^{\beta\gamma}
\chi_L^{i\gamma}
\eeq{7}
\beq
\hp^{i\alpha}_R(x)=e^{i\int_x^{\infty} dz^{\mu}a_{\mu}}
P\left(e^{i\int_x^{\infty} dz^{\mu}B_{\mu}}\right)^{\alpha\beta}
\bar{h}(U^{-1})^{\beta\gamma} \chi_R^{i\gamma}
\eeq{8}

Using the equations of motion for the decoupled fields one can
prove that the fields $\hat{\psi}^i_L$ ($\hat{\psi}^i_R$) are holomorphic
(anti-holomorphic).
In order to show it we analyze in detail
one of them, say  $\hat{\psi}^i_L$,
(the same analysis can be carried out for the other components in
a similar way).

To this end we rewrite eq.\ (\ref{7}) as
\beq
\hp^{i\alpha}_L(z)=e^{\varphi (z)}
{\cal Q}^{\alpha\beta}(z) \chi_L^{i\beta},
\eeq{10}
where we defined (see footnote $2$ for notation)
\beq
\varphi (z) = \phi + i\int_x^{\infty} dz_{\mu} \epsilon_{\mu\nu}
\partial_{\nu} \phi,
\eeq{11}
(note that the field $\eta$ cancels out as it should do, since it
corresponds to the gauge degree of freedom),
and
\beq
{\cal Q}^{\alpha\beta}(z)=
P\left(e^{i\int_x^{\infty} dz^{\mu}B_{\mu}}\right)^{\alpha\gamma}
(V^{-1})^{\gamma\beta}.
\eeq{12}

The field $\varphi(z)$ is the holomorphic component of the free
boson $\phi$. The condition $\bar \partial \varphi =0$
directly follows from the $\phi$ equation of motion.

The holomorphic character of the field ${\cal Q}^{\alpha\beta}(z)$
follows from the equation of motion of $B_{\mu}$, (zero curvature
condition). Calling $\  {\cal U}(x)=
P\left(e^{i\int_x^{\infty} dz^{\mu}B_{\mu}}\right)$ one finds
\beq
i \partial_{\mu} {\cal U}(x) - {\cal U}(x) B_{\mu}(x) =0 .
\eeq{13}
Then,
\beq
i\partial_{\mu}\left( {\cal U} V^{-1}\right) =
{\cal U} \left( B_{\mu}-iV^{-1}\partial_{\mu}V \right)
\eeq{14}
shows that
\beq
{\bar \partial} {\cal Q}=0,
\eeq{15}
while the $z$ derivative can be written as
\beq
\partial {\cal Q}={\cal Q} {\tilde g} \partial {\tilde g}^{-1}
\eeq{15bis}
where ${\tilde g}$ is the field in the (negative level) WZW
sector of the theory (see eq.\ (\ref{a.4})).

Putting all these things together and using the equation of motion
of the free fermion $\bar{\partial} \chi_L=0$, we conclude that
\beq
\bar \partial \hp_L=0 ,~~~~~~i.e.~~\hp_L=\hp_L(z).
\eeq{16}

Similarly the other fields can be written as
\beq
\hp_R(\bar z)=e^{-\bar \varphi(\bar z)}{\bar{\cal Q}}(\bar z) \chi_R(\bar z),
\eeq{17}
where
\beq
\bar \varphi (\bar z) = \phi -i\int_x^{\infty} dz_{\mu} \epsilon_{\mu\nu}
\partial_{\nu} \phi
\eeq{a}
\beq
\bar{\cal Q}(\bar z)=P\left(e^{i\int_x^{\infty} dz^{\mu}B_{\mu}}\right)
U^{-1}.
\eeq{b}
One can easily show that $\bar\varphi$ and $\bar{\cal Q}$ are antiholomorphic
following the same steps as in (\ref{11}-\ref{15}).

The previous heuristic definition of the gauge invariant fermions can be also
motivated by the following natural argument.
One can observe that although  in the decoupling change of variables
eq.\ (\ref{a.2}) the  fields $U$ and $V$ appear, 
only the gauge invariant combination
$\tilde g = V U^{-1}$ is relevant.  Besides, the above semiclassical analysis
shows that we can also factorize the field $\tilde g$ into its holomorphic and
antiholomorphic components as follows
\beq
\tilde g ={\cal Q}^{-1}(z) \bar{\cal Q} (\bar z).
\eeq{qop}

Assuming that this factorization into holomorphic and antiholomorphic
parts is valid at the quantum level (cf.\ \cite{MR,HO}) it is natural
to construct the GIF's with ${\cal Q}$ and $\bar {\cal Q}$ considered as
the holomorphic-antiholomorphic
factors in (\ref{qop}) instead of the (functionals of) the factors 
$U$ and $V$ that
appear in the decoupling change of variables.
Eq.\ (\ref{6}) can then be considered as the  classical counterpart of the
present definition (\ref{10},\ref{17},\ref{qop}).

It is worthwhile to notice that these new fields create physical excitations 
since
they commute with the $SU(k)$ BRST charges, $Q_{BRST}$ and ${\bar Q}_{BRST}$.
The expressions for these charges in the decoupled  picture is \cite{ks}
\beqn
Q_{BRST}=\oint dz :[c^a(z) \left( \chi^{\dagger}_L T^a \chi_L -
\frac{2k+N}{2} \partial {\cal Q}^{-1} {\cal Q}\right) -
1/2 f^{a b c} c^a b^b c^c]:,& &\nonumber\\
{\bar Q}_{BRST}=\oint d{\bar z} :[{\bar c}^a(\bar z) 
\left( \chi^{\dagger}_R T^a \chi_R -
\frac{2k+N}{2} \bar  \partial {\bar {\cal Q}}^{-1} {\bar {\cal Q}}\right) -
1/2 f^{a b c} {\bar c}^a{\bar b}^b {\bar c}^c]:& & 
\eeqn{brst}
where $c, {\bar c}, b, {\bar b}$ are ghost fields.

To show that the gauge invariant fermions defined above
commute with these charges it is sufficient to show that the 
Operator Product Expansion (OPE) with the integrands
in eq.\ (\ref{brst}) is regular. Using the expressions (\ref{10},\ref{17}) 
for the gauge
invariant  fermions, and the transformation
properties of ${\cal Q}$ and ${\bar {\cal Q}}$ under left and 
right $SU(k)$ chiral
rotations, one can show that this is indeed true.

Let us now show how the $SU(N)_k$ WZW fields can be built up from the
gauge invariant fermions.
This can be done by simply taking the product

\beq
\sum_{\alpha=1}^k (\hp_L^{\dagger})^{i\alpha} \hp_R^{j\alpha}=
\sum_{\beta, \gamma =1}^k(\chi_L^{\dagger})^{i\beta} e^{-2\phi}
\left({\cal Q}^{-1}{\bar {\cal Q}}\right)^{\beta\gamma}
\chi_R^{j\gamma}
\eeq{18}
and using the explicit expressions for ${\cal Q}$ as given by
eq.\ (\ref{12}). We finally obtain

\beq
\sum_{\alpha=1}^k (\hp_L^{\dagger})^{i\alpha} \hp_R^{j\alpha}=
\sum_{\alpha=1}^k (\psi_L^{\dagger})^{i\alpha}\psi_R^{j\alpha}=g^{ij}
\eeq{19}
where the last equality follows from eq.\ (\ref{4}).

This simple calculation shows that the gauge invariant fermions,
which create physical states in the holomorphic or
in the antiholomorphic sectors, can be used to construct the
$SU(N)_k$ WZW primary fields $g^{ij}$. This construction exhibits the
holomorphic factorization of the fields $g^{ij}$ \cite{MR,HO}.
The same conclusion applies to the WZW primaries corresponding to
other integrable representations, since they all can be constructed from the
field in the fundamental representation, $g^{ij}$, by taking suitable
symmetrized normal ordered products.

We will now evaluate the OPE between the
energy momentum tensor and the GIF's in order
to prove their primary character and to obtain their conformal dimensions.
The energy momentum tensor can be written as the sum of three independent
contributions, a free fermion part, a free boson part and a WZW part.
Then, the conformal dimensions of the GIF's
are given by the sum of the free fermion contribution, the vertex
operator of the free boson contribution and that of the composite operators
${\cal Q}$.

For the free fermions $\chi_L$ and  $\chi_R$
 the dimensions are $(1/2,0)$ and
$(0,1/2)$ and for the vertex operators $e^{\pm \varphi}$
and $e^{\pm \bar \varphi}$, they are $({-1\over 2Nk},0)$ and
$(0,{-1\over 2Nk})$ respectively.

As for  the conformal dimension of the composites
${\cal Q}(z)$ and ${\bar {\cal Q}}(\bar z)$, they are determined simply
by the conformal transformation properties of the WZW field $\tilde g$.
Since this field is a Virasoro primary of the WZW Conformal Field
Theory and the left and right Virasoro algebras commute with each
other, the holomorphic and antiholomorphic operators
 ${\cal Q}(z)$ and ${\bar {\cal Q}}(\bar z)$
are primaries of the left and right Virasoro algebras separately.

In fact,  ${\cal Q}$ transforms in the fundamental representation
of the affine Lie algebra $SU(k)_{-N-2k}$, {\it i.e.,} its OPE
with the affine current is given by

\beq
{\tilde J}^a(z){\cal Q}(w)=\frac{t^a{\cal Q}(w)}{z-w}+
{\cal N}[{\tilde J}^a{\cal Q}](w)+r.t.,
\eeq{20}
where ${\cal N}$ denotes normal ordering and $r.t.$ stands for
regular terms.

Using this equation and the Sugawara representation of the energy momentum
tensor \cite{KZ}, we obtain:

\beq
{\tilde T}(z){\cal Q}(w)=-\frac{k^2-1}{2k(k+N)}\frac{{\cal Q}(w)}{(z-w)^2}+
\frac{2}{z-w}\partial_w{\cal Q}(w)+r.t.,
\eeq{21}
from which one can read off the dimension  of ${\cal Q}$,
$h_{\cal Q}=-\frac{k^2-1}{2k(k+N)}$.

Adding to $h_{\cal Q}$ the contributions from the free fermions and that
of the vertex operator of the free boson, we get for the dimension
of the GIF operator the expression
\beq
h_{\hp}=-\frac{k^2-1}{2k(k+N)} +\frac{1}{2} - \frac{1}{2Nk}=
\frac{N^2-1}{2N(k+N)}.
\eeq{24}

This shows that the GIF's
have the conformal dimensions corresponding to the
holomorphic or antiholomorphic factors of the $SU(N)_k$ WZW primaries
(\ref{4}).

According to \cite{TK} this means that the GIF's are the vertex operators
of the WZW theory. Usually the so-called Chiral Vertex Operators are
introduced in this context \cite{TK,MS} and can be constructed as
appropriate projections of the GIF's.
They formally correspond to (considering for simplicity the case of $N=2$)
\beq
\Phi\CVO{j}{i}{k}=\Pi_i {\cal S}\left(\hp \hp .. \hp\right) \Pi_k ~ ,
\eeq{WW}
where $\Pi_i$ stands for the projector on the integrable representation
of spin $i$ and $\cal S$ is the Young symmetrizer that projects the
product of $2j$ GIF's onto the representation of spin $j$.

\vspace{.5cm}

\noindent {\it iv) Braiding relations among GIF's}

\vspace{.5cm}

We now evaluate the OPA satisfied by the GIF's,
which can be easily
calculated using the explicit expressions in eqs.\ (\ref{10},\ref{17}).

Let us consider first the OPE of two fields $\hp^{i\alpha}_L$:
\beq
\hp^{i\alpha}_L(z) \hp^{j\beta}_L(w)
=e^{\varphi (z)}
{\cal Q}^{\alpha\alpha'}(z) \chi_L^{i\alpha'}(z)
e^{\varphi (w)}
{\cal Q}^{\beta\beta'}(w) \chi_L^{j\beta'}(w).
\eeq{25}

The OPE of the two $U(1)$ bosonic vertex operators appearing in (\ref{25}) is
simply given by
\beq
e^{\varphi (z)}e^{\varphi (w)}=
\frac{1}{(z-w)^{\frac{1}{N-k}}}e^{2\varphi (w)} +...~,
\eeq{26}
and that of the free fermions, which is conveniently separated 
into symmetric and
antisymmetric combinations, by
\beqn
\chi^{i\alpha}_L(z)\chi^{j\beta}_L(w)=
1/2( :\chi^{i\alpha}_L(w)\chi^{j\beta}_L(w):-
:\chi^{j\alpha}_L(w)\chi^{i\beta}_L(w):)+\nonumber\\
1/2(
:\chi^{i\alpha}_L(w)\chi^{j\beta}_L(w):+
:\chi^{j\alpha}_L(w)\chi^{i\beta}_L(w):)+...~.
\eeqn{27}

Finally for the OPE of the WZW vertex operators we have
\beq
{\cal Q}^{\alpha\alpha'}(z){\cal Q}^{\beta\beta'}(w)=
\sum_l {\cal C}_{{\cal Q}{\cal Q}}^{(l)} z^{h_l-2h_{\cal Q}}\left[
\Phi_l^{\alpha \beta \alpha'\beta'}(w)+...\right]
\eeq{28}
where the $\Phi_l$'s are the primary fields associated with the integrable
representations of the affine chiral algebra  \cite{KZ}, and the dots stand for
their descendants.
In our case, we have
\beq
{\cal Q}^{\alpha\alpha'}(z){\cal Q}^{\beta\beta'}(w)=
\frac{{\cal C}^{A}_{{\cal Q}{\cal Q}}}{(z-w)^{-h_A+2h_{\cal Q}}}
\Phi_A^{\alpha\beta\alpha'\beta'}(w)+
\frac{{\cal C}^{S}_{{\cal Q}{\cal Q}}}{(z-w)^{-h_S+2h_{\cal Q}}}
\Phi_S^{\alpha\beta\alpha'\beta'}(w)+...
\eeq{29}
where $\Phi_A$ and $\Phi_S$ are the antisymmetric and symmetric
channels of the product in the right hand side of (\ref{28})
with dimensions $h_A=\frac{(2-k)(k+1)}{k(k+N)}$,
$h_S=\frac{(k+2)(1-k)}{k(k+N)}$ respectively.

Combining eqs.\ (\ref{25},\ref{26},\ref{29}) we obtain 
\beq
\hp^{i\alpha}_L(z) \hp^{j\beta}_L(w)= (z-w)^{-h_{\cal S}} {\cal
S}^{ij}_{\alpha\beta}(w)+ (z-w)^{-h_{\cal A}} {\cal
A}^{ij}_{\alpha\beta}(w)+...~,
\eeq{30} 
where $ {\cal S}^{ij}_{\alpha\beta}$  ($ {\cal A}^{ij}_{\alpha\beta}$) is
symmetric (antisymmetric) in the indices $\alpha, \beta$ and
antisymmetric (symmetric) in the indices $i, j$. Their conformal
weights $h_{\cal S}$ and $h_{\cal A}$ are given respectively by:

\beq 
h_{\cal S}=\frac{1+N}{N (N+k)} , \ \ \ \ \ {\rm and}\ \ \ \ \
\ h_{\cal A}=\frac{1-N}{N (N+k)}~. 
\eeq{31}

As already stressed, the Chiral Vertex Operators
$\Phi\CVO{i}{j}{k}$ introduced in \cite{TK,MS} correspond to
suitable projections of the GIF's over the integrable
representations of the $SU(N)_k$ affine algebra (see eq.\
(\ref{WW})). With these operators one can verify explicitly the
$N=2$ braiding relations \cite{TK,MS} 
\beq
\Phi\CVO{k_1}{j_1}{p}(z_1) \Phi\CVO{k_2}{p}{j_2}(z_2) = \sum_{p'}
B_{pp'}\! \left[ \matrix{ k_1 & k_2 \cr j_1 & j_2 \cr} \right]\!
\Phi\CVO{k_2}{j_1}{p'}(z_2) \Phi\CVO{k_1}{p'}{j_2}(z_1)~. 
\eeq{ZZ}
\ \indent From this we can conclude that the GIF's can be considered as
quasiparticle operators with non-Abelian braiding in the sense of
\cite{MR,BW}. Correlators of these operators can be computed
following the same steps as in \cite{KZ}.

\vspace{.5cm}

\noindent {\it v) Conclusions}

\vspace{.5cm}

We have shown in this paper that the fermionic coset representation
of the $SU(N)_k$ WZW theory allows for the explicit construction of
the vertex operators \cite{TK}, fundamental physical fields that
can be used to build up the WZW primaries.
These gauge invariant fermion fields are holomorphic and transform as
the holomorphic part of the primaries under the action of
the full chiral algebra.

Moreover, they create physical excitations and their modes
can then be used to construct the physical Hilbert space,
which will consist of states of left and right moving
quasiparticles.
Due to the braiding relations satisfied by the GIF's,
they can be interpreted as quasiparticle operators with
non-Abelian statistics according to \cite{MR,BW}.

The connection between the Fock space obtained through this
procedure could be compared with the spinon Fock space
constructed in \cite{BLS}, to understand the relation of the
quasiparticles created with the modes of the GIF's 
and the non-Abelian spinons.

\vspace{.5cm}

\noindent {\it  Acknowledgements}: D.C.C. would like to thank A.\ Honecker
for helpful discussions and Fundaci\'on Antorchas for partial support.
E.F.M. is partially supported by CUNY Collaborative Incentive Grant
991999. G.L.R. thanks Prof. R. Jackiw for kind hospitality at CTP, MIT.
G.L.R. was supported in part by funds provided by the U.S. Department of
Energy
(D.O.E.) under cooperative research agreement No. DF-FC02-94ER40818.

\end{document}